\documentclass[a4paper,10.5pt,twoside]{article}
\usepackage[top=3cm, bottom=3cm, inner=3cm, outer=3cm, includehead]{geometry}
\usepackage{fancyhdr}
\usepackage{xurl}
\usepackage{graphicx}
\usepackage{alltt}
\usepackage{amsmath}
\usepackage[hidelinks, pdftex]{hyperref}
\usepackage[T1]{fontenc}
\usepackage[utf8]{inputenc}
\usepackage{lmodern}
\usepackage{csquotes}
\usepackage[symbol]{footmisc}

\usepackage[
backend=biber,
style=numeric,
sorting=ynt
]{biblatex}
\usepackage[english]{babel}

\begin{document}
\thispagestyle{empty}
\fancyhead[LE]{\thepage\ \ \ \ Towards End-to-End Model-Agnostic Explanations for RAG Systems}
\fancyhead[RO]{Sudhi et al.\ \ \ \ \thepage}
\begin{center}
\LARGE
\textbf{Towards End-to-End Model-Agnostic Explanations for RAG Systems}\\[12pt]
\normalsize
\textbf {
Viju Sudhi\footnotemark[1],
Sinchana Ramakanth Bhat\footnotemark[2], 
Max Rudat\footnotemark[2], 
Roman Teucher\footnotemark[2] and 
Nicolas Flores-Herr\footnotemark[2]
}\\[4pt]
\end{center}

\footnotetext[1]{Correspondence to viju.sudhi@uni-bielefeld.de }
\footnotetext[1]{Affiliation: Bielefeld University (Work done while the author was affiliated with Fraunhofer IAIS)}
\footnotetext[2]{Affiliation: Fraunhofer IAIS}

\begin{abstract}
\normalsize
Retrieval Augmented Generation (RAG) systems, despite their growing popularity for enhancing model response reliability, often struggle with trustworthiness and explainability. In this work, we present a novel, holistic, model-agnostic, post-hoc explanation framework leveraging perturbation-based techniques to explain the retrieval and generation processes in a RAG system. We propose different strategies to evaluate these explanations and discuss the sufficiency of model-agnostic explanations in RAG systems. With this work, we further aim to catalyze a collaborative effort to build reliable and explainable RAG systems.

\end{abstract}

\section{Introduction}


RAG systems aim at improving response generation of Large Language Models (LLMs) \cite{lewis2020retrieval, gao2023retrieval, gupta2024comprehensive}. 
A typical RAG system is composed of a retriever in conjunction with a generator. Given a user question $q$, the retriever from a collection of documents returns the most relevant documents $d_{i}$. These documents together with an instruction compose the prompt $x$ which is then fed to the LLM-based generator. The generator finally returns a response $y$ to the user - more reliable than the one it generates from its model weights alone. 
However, since the models used are not intrinsically explainable, end-users often find such RAG systems less trustworthy \cite{zhou2024trustworthiness}. To mitigate this, in this work\footnote[4]{Code: \url{https://github.com/fraunhofer-iais/explainable-lms}}, we borrow ideas presented in our earlier works attempting to individually explain the retriever \cite{bite-rex} and the generator \cite{rag-ex}; and combine these strategies to build a holistic end-to-end framework towards model agnostic explanations for RAG systems. Our framework can explain retrievers (utilizing dense embedding models) and generators (open-source or proprietary) in an open-book QA setup. We present the framework as a "one-fits-all" solution considering the plethora of emerging embedding and generator models.

\begin{figure*}[t!]
    \centering
    \includegraphics[width=\textwidth]{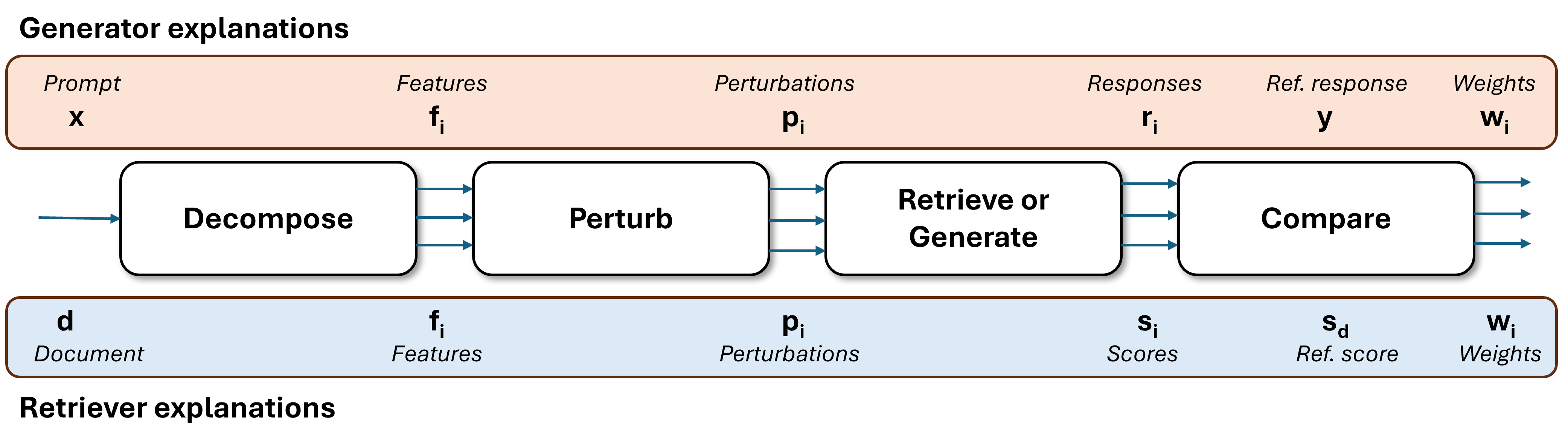}
    \caption{Overview of the explanation framework. For retriever explanations, the retrieved document $d$ is fed into the explainer. This is then decomposed to obtain features $f_i$ and perturbed to obtain perturbations $p_i$. The similarity between the perturbed document $p_i$ and the user question is computed as $s_i$. This is compared against the reference score $s_d$ finally resulting in feature importance weight $w_i$. For generator explanations, the input to the explainer is the prompt $x$. This is decomposed to obtain $f_i$ and based on each feature, the input is perturbed to obtain $p_i$. For each perturbed input, the generator generates a response $r_i$ which is then compared against the reference response $y$ leading to the feature importance weight $w_i$.}
    \label{fig:blocks}
\end{figure*}

\section{Methodology}

\begin{figure*}[t!]
    \centering
    \includegraphics[width=\textwidth]{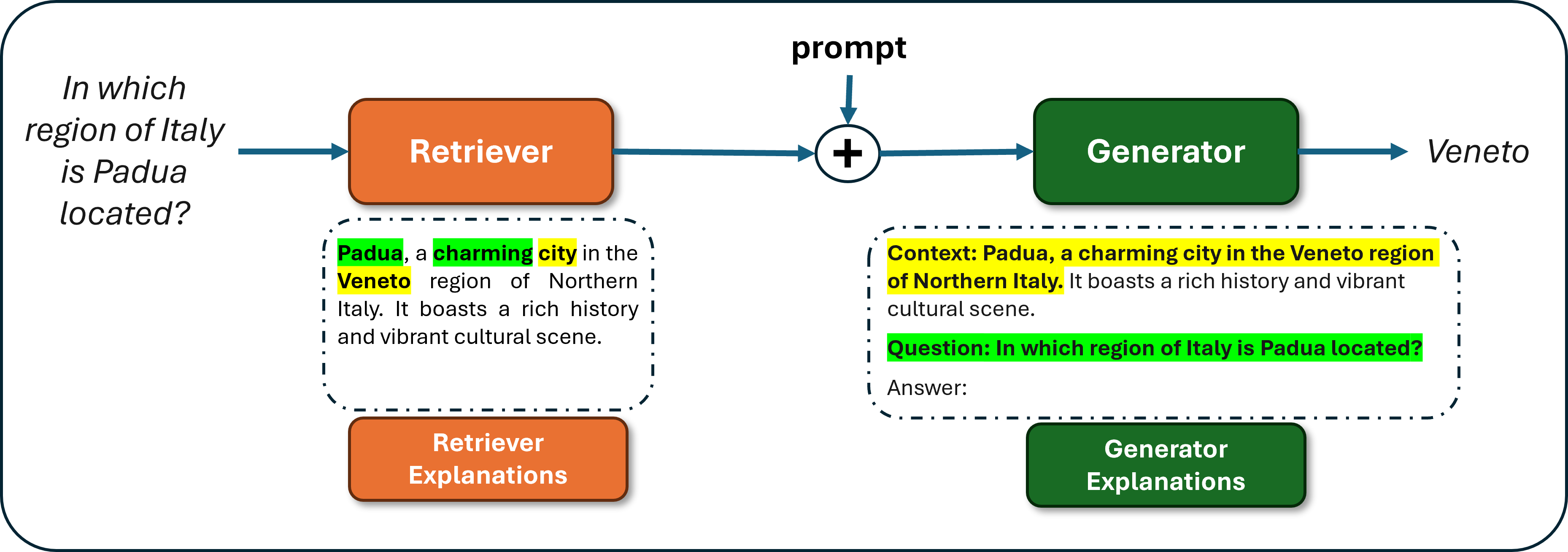}
    \caption{An exemplary visualization of the explanation framework.}
    \label{fig:example}
\end{figure*}

In the proposed end-to-end explanation framework, we aim at: \textbf{(i) explaining the retrieval process} to answer why the retriever retrieved $d$ given $q$, and \textbf{(ii) explaining the generation process} to answer why the generator generated $y$ given $x$. While the former helps the user understand the contribution of individual document terms, the latter helps them to understand which parts of the input the LLM focused on to generate the final answer. By presenting component-wise explanations, we allow the user to be a better judge in carefully choosing different retriever and generator models to finally compose their reliable and explainable RAG system. An exemplary visualization is presented in \ref{fig:example}.

The explainer starts with deciding what should be explained: $d_i$ in the case of the retriever and $x$ in the case of the generator. As illustrated in Figure \ref{fig:blocks}, we then employ the same set of components as outlined below to explain the retrieval and generation processes. 
\begin{itemize}
    \item \textbf{Decompose:} Firstly, the inputs ($d$ or $x$) are decomposed to the features $f_i$ according to the preferred granularity. For retriever explanations, we use \textit{word-level} granularity to study the significance of individual terms in the document $d$. The generator explanations should shed light into the different parts of the input prompt $x$ and therefore, we use \textit{sentence-level} granularity for generator explanations.
    \item \textbf{Perturb:} Based on each of the decomposed features $f_i$, the inputs ($d$ or $x$) are perturbed to yield the perturbations $p_i$. We examined different perturbation strategies\footnote{We advise the readers to find more details about the strategies in the work \cite{rag-ex}.} and observed that the simplest \textit{leave one feature out} strategy (where each feature in the input is individually left out) yields the most intuitive explanations.
    \item \textbf{Retrieve or Generate:} The perturbed inputs $p_i$ are then fed to the corresponding component. For retriever explanations, we (i) first, utilize the base retriever which yielded the relevant document to now embed the perturbed inputs and (ii) then, compute the cosine similarity score against the user question $s_i$. For generator explanations, we (i) first, feed the perturbed inputs to the base generator and (ii) then, generate responses $r_i$ for each $p_i$.
    \item \textbf{Compare:} The resulting retriever scores $s_i$ and the generator responses $r_i$ are compared against the reference retriever score $s_d$ and the reference generator response $y$ respectively to study the relative similarities (of the scores and the texts). We quantify the importance of the feature $f_i$ to the similarity scores further normalized and negated to obtain the dissimilarity scores denoted as $w_i$. These weights indicate the feature importance for the corresponding process.
\end{itemize}

The components summarized above are agnostic to the language and models used, thereby allowing the usage of any retriever or generator model. In our open-source user interface, the features are marked with different color scales to help users easily identify and distinguish the most important features from the others.

\section{Evaluation}
We evaluated the explanations from our framework considering different facets and report our results briefly as follows. We advise the readers to refer to our work BiTe-REx \cite{bite-rex} and RAG-Ex \cite{rag-ex} for more details on evaluation methods and metrics.

\paragraph{Intuitiveness of the explanations} As the first step, we wanted to assess how intuitive the end-users find the explanations from the framework. We designed user studies in which we instructed users to annotate features that they believed were significant for retrieval or generation. Upon comparing these features against the ones the explainer yielded as significant for the process, we were able to understand how similar the user and the framework "explain" the processes. Our retriever explanations were found to be 64.7\% \textit{complete} and the generator explanations yielded an F1 score of 76. 9\% against the end-user annotations. 

\paragraph{Co-relation with downstream task model performance} 
We observed that the models' downstream task performance was consistently higher when the generator explainer yielded more intuitive features as the significant ones. In other words, if the model \textit{looked} into the most relevant parts of the context - in this use case, the question itself and the part of the context where the potential answer lies, its chances to generate the correct answer are higher.

\paragraph{Sufficiency of model-agnostic explanations} We also studied how users received the model-agnostic generator explanations against the available model-intrinsic approaches. They rated our framework to be 3.42 \textit{complete} and 3.45 \textit{correct} against 3.98 and 4.04, respectively, for the compared model-intrinsic approach. Despite the gap in these ratings, we strongly advocate model-agnostic explanations for RAG systems, given the flexibility it offers by accommodating both open-source and proprietary models. 

\section{Limitations}
We acknowledge that the framework is limited in its utility considering use cases other than RAG systems. We choose \textit{word-level} granularity for explaining the retrieval process overlooking how dense embedding models are otherwise trained. Unlike model-intrinsic approaches which aim at understanding the technical workings of the language models in general, model-agnostic approaches like ours tend to provide \textit{approximated} explanations to the user. This caters rather better to the end-users of the system, but does not necessarily expose the working of the model by itself.

\section{Conclusion}
We present an end-to-end model agnostic framework to explain the retrieval and generation processes in RAG systems. As future work, we plan to extend the evaluation of our framework by extending our qualitative analysis to evaluate intuitiveness and satisfaction; asking participants to rate explanations based on clarity, relevance, and trustworthiness. We also aim to investigate better quantitative measures for evaluating explanations.




\printbibliography

\end{document}